\documentclass{aastex}
\usepackage{emulateapj5}
\usepackage{apjfonts}
\usepackage{epsf}

\newcommand{\HA}{\ensuremath{\mathrm{H}\alpha}}
\newcommand{\HB}{\ensuremath{\mathrm{H}\beta}}
\newcommand{\OII}{[O {\sc ii}]$\lambda3727$}
\newcommand{\OIII}{[O {\sc iii}]$\lambda\lambda$4959,5007}
\newcommand{\NII}{[N {\sc ii}]$\lambda\lambda6548,6584$}

\begin{document}

\title{THE H$\alpha$ LUMINOSITY FUNCTION AND STAR FORMATION RATE AT $z
\approx 0.24$ BASED ON SUBARU DEEP IMAGING DATA\altaffilmark{1}}

\author{Shinobu S. Fujita   \altaffilmark{2},
        Masaru Ajiki        \altaffilmark{2},
        Yasuhiro Shioya     \altaffilmark{2},
        Tohru Nagao         \altaffilmark{2},
        Takashi Murayama    \altaffilmark{2},
        Yoshiaki Taniguchi  \altaffilmark{2},
		Kazuyoshi Umeda		\altaffilmark{2},
		Sanae Yamada		\altaffilmark{2},
		Masafumi Yagi       \altaffilmark{3},
		Sadanori Okamura    \altaffilmark{4,5}, and 
		Yutaka Komiyama     \altaffilmark{6}
		}

\altaffiltext{1}{Based on data collected at 
	Subaru Telescope, which is operated by 
	the National Astronomical Observatory of Japan.}
\altaffiltext{2}{Astronomical Institute, Graduate School of Science,
        Tohoku University, Aramaki, Aoba, Sendai 980-8578, Japan}
\altaffiltext{3}{National Astronomical Observatory, 
	Mitaka, Tokyo 181-8588, Japan}
\altaffiltext{4}{Department of Astronomy, Graduate School of Science,
        University of Tokyo, Tokyo 113-0033, Japan}
\altaffiltext{5}{Research Center for the Early Universe, School of Science,
        University of Tokyo, Tokyo 113-0033, Japan}
\altaffiltext{6}{Subaru Telescope, National Astronomical Observatory, 
	650 N.A'ohoku Place, Hilo, HI 96720}

\shortauthors{Fujita et al.}
\shorttitle{H$\alpha$ luminosity functions at $z \approx 0.24$}

\begin{abstract}
We have carried out a deep imaging survey for H$\alpha$ emitting galaxies
at $z\approx$0.24 using a narrowband filter tuned with the redshifted
line. The total sky area covered is 706 arcmin$^2$ within a redshift
range from 0.234 to 0.252 ($\delta z$=0.018).
This corresponds to a volume of 3.9$\times10^3$  Mpc$^3$ when 
$\Omega_{\rm matter}=0.3$, $\Omega_\Lambda=0.7$, and $H_{\mathrm{0}}$=70
km s$^{-1}$ Mpc$^{-1}$ are adopted.
We obtain a sample of 348 H$\alpha$ emitting galaxies whose observed
emission-line equivalent widths are greater than 12 \AA. We find an 
extinction-corrected H$\alpha$ luminosity density of  
$10^{39.65^{+0.08}_{-0.12}}$ ergs s$^{-1}$ Mpc$^{-3}$.
Using the Kennicutt relation between the H$\alpha$ luminosity and star 
formation rate, the star formation rate density in the covered volume is estimated as 
$0.036^{+0.006}_{-0.012}$ $M_{\sun}$ yr$^{-1}$ Mpc$^{-3}$.
This value is higher by a factor of 3  than the local SFR density.
\end{abstract}

%\maketitle

\section{INTRODUCTION}

The star formation rate density is one of the important observables for
our understanding of galaxy formation and evolution. Madau et al. (1996)
compiled the evolution of star formation rate (SFR) density,
$\rho_{\rm SFR}$, as a function
of redshift from the local universe to high-redshift. 
It is known that $\rho_{\rm SFR}$  steeply increases from $z \simeq 0$ to $z\sim1$
(Gallego et al. 1995; Tresse \& Maddox 1998, hereafter TM98;
Songaila et al. 1994; Ellis et al. 1996; Lilly et al. 1996; Hogg et al. 1998).
However, $\rho_{\rm SFR}$ appears to be flat beyond $z=2$  (Madau, Pozzetti, \&
Dickinson 1998; Pettini et al 1998; Steidel et al. 1999; Barger et al. 2000;
Fujita et al. 2003a; see also Trentham, Blain, \& Goldader 1999). 

Although recent investigations have been devoted to new estimates of
$\rho_{\rm SFR}$ at high redshift, it seems also important to provide
new estimates in nearby universe because only several investigations 
have been available between $z \simeq 0$ to $z \sim 1$. 
For example, $\rho_{\rm SFR}$ at $z \simeq 0.2-0.3$ is based solely on 
the observation by TM98.
In this Letter, we present a new estimate of $\rho_{\rm SFR}$ at
$z \approx 0.24$ based on the largest sample (more than 300 galaxies)
studied so far, which was obtained with Suprime-Cam on the Subaru Telescope.

Throughout this paper, magnitudes are given in the AB system.
We adopt a flat universe with $\Omega_{\rm matter} = 0.3$, 
$\Omega_{\Lambda} = 0.7$,
and $h=0.7$ where $h = H_0/($100 km s$^{-1}$ Mpc$^{-1}$).

\section{OBSERVATIONS AND DATA REDUCTION}

We have made a very deep optical imaging survey of H$\alpha$ emitters 
at $z \approx 0.24$ in a sky area surrounding the SDSSp J104433.04$-$012502.2 
at redshift of 5.74
(Fan et al. 2000; Djorgovski et al. 2001; Goodrich et al. 2001)
using the Suprime-Cam (Miyazaki et al. 2002) on 
the 8.2 m Subaru telescope (Kaifu 1998) at Mauna Kea Observatories.
The Suprime-Cam consists of ten 2k$\times$4k CCD chips and 
provides a very wide field of view; $34^\prime \times 27^\prime$.

In this survey\footnote{This survey is originally intended to search for
Ly$\alpha$ emitters
at $z \approx 5.7$ (Ajiki et al. 2002; Shioya et al. 2002).},
we used the narrow-passband filter, {\it NB}816, centered at
8150 \AA ~ with the passband of $\Delta\lambda$(FWHM) = 120 \AA;
the central wavelength corresponds to a redshift of 0.24 for 
H$\alpha$ emission. 
We also used broad-passband filters,
$B$, $R_{\rm C}$, $I_{\rm C}$, and $z'$. 
All the observations were done under
photometric condition and the seeing size was between 0.7 arcsec
and 1.3 arcsec during the run.

The individual CCD data were reduced and
combined using IRAF and the mosaic-CCD data reduction software 
developed by Yagi et al. (2002). Detail of flux calibration will
be presented in Fujita et al. (2003b).
The combined images for the individual bands were aligned and 
smoothed with Gaussian kernels to match their seeing sizes. 
The region contaminated by fringing pattern is masked.
The final images cover a contiguous 706 arcmin$^2$ area
with a PSF FWHM of $1.''4$. A total volume of $3.9\times10^3$Mpc$^3$
is probed in our {\it NB}816 image.
The net exposure times of the final images are 28, 80, 56, 86, and 600
minutes for $B$, $R_{\rm C}$, $I_{\rm C}$, $z'$, and {\it NB}816, respectively.
The limiting magnitudes are $B=26.6$, $R_{\rm C}=26.2$, $I_{\rm C}=25.9$,
$z'=25.3$, and {\it NB}816$=26.0$ for a $3\sigma$ detection on a $2''.8$
diameter aperture.

Catalogues of the objects are made using SExtractor (Bertin \& Arnouts 1996).
The objects are detected using the \emph{double-image mode}:
the narrowband frame is used as a reference image for detection and 
then the flux is summed up in 14-pixel diameter apertures in 
all images. This aperture size is $2''.8$, i.e.,  2$\times$FWHM of the stellar
objects in the final image of $R_{\rm C}$-band.

\section{RESULTS}

\subsection{Selection of {\it NB}816-Excess Objects}

\begin{figure}
\epsscale{0.5}
\plotone{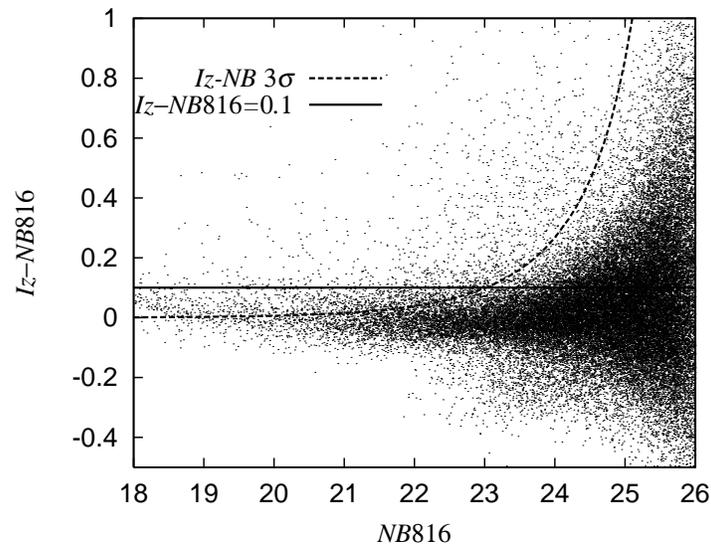}
\caption{Objects detected down to $NB816=26$ in the $NB816$-selected catalog.
The horizontal solid line corresponds to $Iz-NB816=0.1$. 
Dashed line shows the distribution of $3\sigma$ error. Because bright sources
are saturated in the continuum frame, {\it Iz}-{\it NB}816 is not zero at
{\it NB}816$\lesssim19$.
\label{Ha:Iz-NBvsNB}}
\end{figure}

Since the central wavelength of the $I_{\rm C}$ is bluer than that of the NB816
filter, we calculated magnitude that we refer to as the ``$Iz$
continuum'', using a linear combination ($Iz = 0.76I_{\rm C}+0.24
z^{\prime}$) of the $I_{\rm C}$ and $z'$ flux densities; a 3 $\sigma$
photometric limit of $Iz\simeq26.0$ in a $2''.8$ diameter aperture.
This enables us to more precisely sample the continuum at the same effective
wavelength as that of the {\it NB}816 filter. Following the manner of the previous surveys using
narrowband filter and taking the scatter in the $Iz$-{\it NB}816
color and our survey depth into account, candidate line-emitting objects are selected with the
criterion of $Iz$-{\it NB}816 $\geq$ max(0.1, $3 \sigma$ error of $Iz$-{\it NB}816);
note that $Iz$-{\it NB}816 = 0.1 corresponds to $EW_{\rm obs} \approx12$ \AA.
We compute the $3\sigma$ of the color as
$3\sigma_{Iz-\mathit{NB}816}=-2.5\log
(1-\sqrt{(f_{3\sigma_{\mathit{NB}816}})^2+(f_{3\sigma_{Iz}})^2}/f_{\mathit{NB}816})$.
This criterion is shown by the solid and dashed lines in Figure \ref{Ha:Iz-NBvsNB}.
There are 1224 sources that satisfy the above criterion. They are brighter than
the limiting magnitude at each band.

\subsection{Selection of {\it NB}816-Excess Objects at $z\approx0.24$}

A narrowband survey of emission line galaxies can potentially 
detect galaxies with  different emission lines
at different redshifts. If the source redshift and the rest frame wavelength
of the line act to place it inside the narrowband filter, the line
will be detected if it is sufficiently strong.
The emission lines we would expect to detect are H$\alpha$, H$\beta$, 
\OIII, and \OII ~(Tresse et al. 1999; Kennicutt 1992b)
as the narrowband filter passband is too wide to
separate \NII\ from \HA. In Table~\ref{Ha:tab:cover} we show
the different redshift coverage for each line.

\begin{figure}
\epsscale{0.5}
\plotone{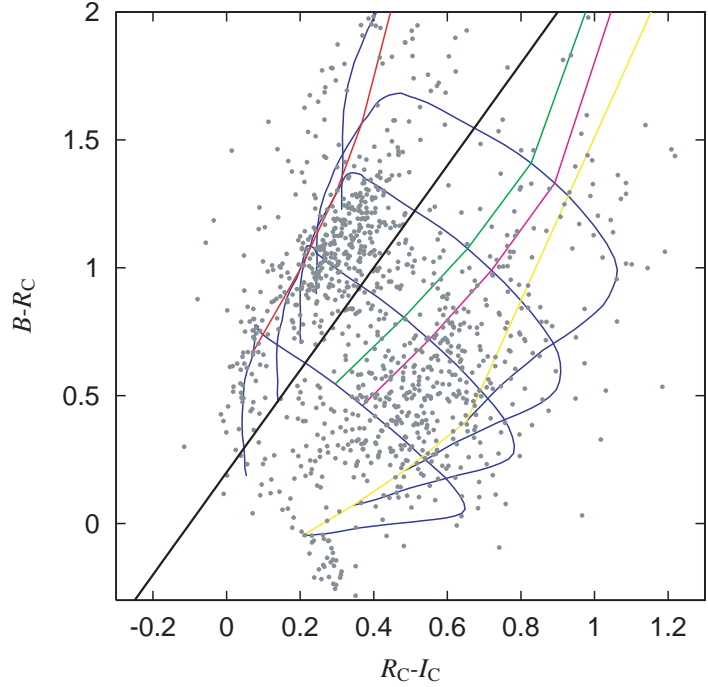}
\caption{Plot of $B-R_{\rm C}$ vs. $R_{\rm C}-I_{\rm C}$ for the 1224 sources found with
emitter selection criteria (gray circles). Colors of model galaxies (SB, Irr, Scd, Sbc, and E)
from $z=0$ to $z=1.2$ are shown with blue lines. Colors of $z=0.24$, $0.64$, $0.68$, and $1.18$
(for H$\alpha$, [O {\sc iii}], H$\beta$, and [O {\sc ii}] emitters, respectively) are shown
with red, green, purple, and yellow lines. We select the sources above the black line as H$\alpha$
emitters.
\label{Ha:BRIcolor}}
\end{figure}

In order to distinguish H$\alpha$ emitters at $z \approx 0.24$ from
emission-line objects at other redshifts,
we investigate their broad-band color properties comparing 
observed colors of our 1224 emitters with model ones that are estimated by using 
the  population synthesis model GISSEL96 (Bruzual \& Charlot 1993).
In Figure \ref{Ha:BRIcolor}, we show the $B-R_{\rm C}$ vs. $R_{\rm C}-I_{\rm C}$
color-color diagram of the 1224 sources and the loci of model galaxies.
Then we find that H$\alpha$ emitters at $z\approx0.24$ can be selected by
adopting the following three criteria (see for detail, Fujita et al. 2003b);
(1) $B-R_{\rm C} > 2(R_{\rm C} - I_{\rm C}) + 0.2$,
(2) $z^{\prime} - \mathit{NB}816 > 0$, and
(3) $I_{\rm C} - \mathit{NB}816 < 0.75(z^{\prime}-\mathit{NB}816) + 0.35$.
These criteria give us a sample of 348 \HA\ emitting galaxy candidates.
In particular, the first color criterion is a quite nice discriminator
between H$\alpha$ emitters at $z \approx$ 0.24 and other emitters at 
different redshifts and thus it is expected that there is few contamination
in our sample.

We adopt the same method as that used by Pascual et al. (2001) to calculate \HA\ equivalent width.
The flux density in each filter can be expressed as the sum of the line
flux and the continuum flux density (the line is covered by both filters):
\begin{equation}
	f_{\mathit{NB}} = f_C + \frac{F_L}{\Delta \mathit{NB}}\textrm{, and }
	f_{Iz} = f_C + 0.76\frac{F_L}{\Delta I}
\end{equation}
with $f_C$ the continuum flux density; $F_L$ the line flux; $\Delta NB$ and
$\Delta I$ the narrow and $I_{\rm C}$ band filter effective widths and $f_{NB}$
and $f_{Iz}$ the flux density in each filter. Then the line flux, continuum
flux density, and equivalent width can be expressed as follows:
\begin{eqnarray}
	F_L &=& \Delta \mathit{NB}\frac{f_{\mathit{NB}}-f_{Iz}}{1-0.76(\Delta \mathit{NB}/\Delta I)}, \\
	f_C &=& \frac{f_{Iz}-0.76f_{\mathit{NB}}(\Delta \mathit{NB}/\Delta I)}{1-0.76(\Delta \mathit{NB}/\Delta I)},\\
	EW_{\rm obs} &=& \frac{F_L}{f_C} = \Delta \mathit{NB}
		\left(\frac{f_{\mathit{NB}}-f_{Iz}}{f_{Iz}-0.76f_{\mathit{NB}}(\Delta \mathit{NB}/\Delta I)}\right),
\end{eqnarray}
where $\Delta I=1269$ \AA\ for the $I_{\rm C}$ band and $\Delta\mathit{NB}=120$ \AA\ for the $\mathit{NB}816$ band.
The minimum line flux in our 348 sources is $1.2\times 10^{-17}$ ergs s$^{-1}$ and the minimum continuum
flux density reaches $3.0\times10^{-30}$ ergs s$^{-1}$ Hz$^{-1}$ ($\approx25.2$ in AB magnitude).

\subsection{H$\alpha$ Luminosity}

In order to obtain the H$\alpha$ luminosity for each source, 
we correct for the presence of \NII\ lines.
Further, we also apply a mean internal extinction
correction to each object. For these two corrections,
we have adopted the flux ratio of
$f(\HA)/f($\NII$) = 2.3$ (obtained by Kennicutt 1992a; Gallego et al. 1997; used by TM98;
Yan et al. 1999; Iwamuro et al. 2000) and $A_{\HA}=1$.

We also apply a statistical correction (28\%; the average value of flux decrease
due to the filter transmission) to measured flux because the filter transmission
function is not square in shape.
The \HA\ flux is given by:
\begin{eqnarray}
	f_{\rm cor}({\rm H}\alpha) = f({\rm H}\alpha&+&[\textrm{ N {\sc ii}}])\times
		\frac{f({\rm H}\alpha)}{f({\rm H}\alpha)+f([\textrm{ N {\sc ii}}])} \nonumber \\
		&\times&10^{0.4A_{{\rm H}\alpha}}\times 1.28.
\end{eqnarray}
We calculate $f({\rm H}\alpha+[\textrm{ N {\sc ii}}])$ from the total magnitude;
note that the aperture size is determined from the $I_{\rm C}$-band image.
Finally the \HA\ luminosity is given by $L({\rm H}\alpha) = 4\pi d_{\rm L}^2f_{\rm cor}({\rm H}\alpha)$
using the redshift of the line at the center of the filter $z=0.242$ and
the luminosity distance, $d_{\rm L}$.

\section{LUMINOSITY FUNCTION AND STAR FORMATION RATE}

\subsection{Luminosity function of \HA\ emitters}

\begin{figure}
\epsscale{0.5}
\plotone{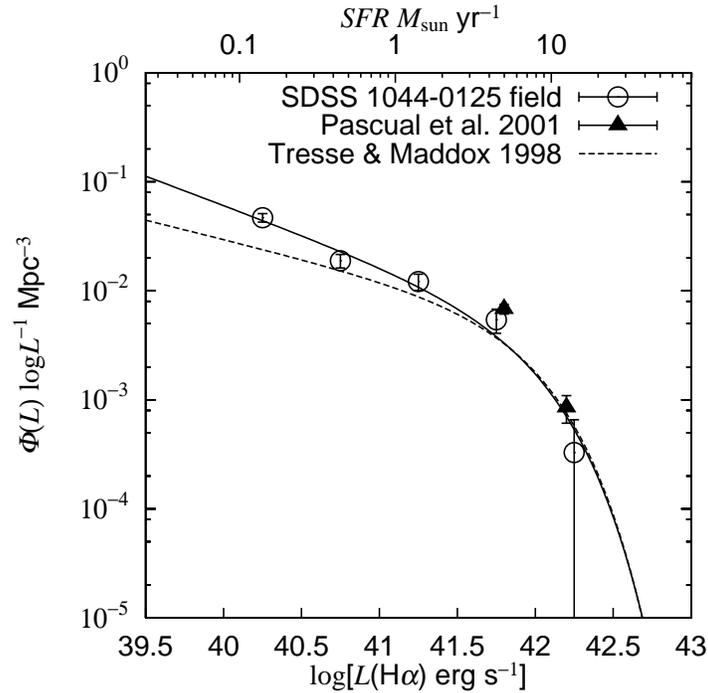}
\caption{Luminosity function at $z\approx0.24$ (open circles). Solid line is the best fitted Schechter function.
The data points with $L$(\HA)$\geq 10^{40}$ ergs s$^{-1}$ are used in the LF fitting.
The TM98 H$\alpha$ luminosity function at $z\leq0.3$ is shown with dashed line.
The data points of Pascual et al. (2001) are also shown by filled triangles.
\label{Ha:LF}}
\end{figure}

In order to investigate the star formation activity in galaxies at $z \approx 0.24$,
we construct the \HA\ luminosity function (LF) for our H$\alpha$ emitter sample
by using the relation, 
$\Phi(\log L(\HA)) \Delta\log L(\HA) = \sum_i 1/V^i$,
where $V^i$ is the comoving volume, and the sum is over galaxies with \HA\
luminosity within the interval [$\log L(\HA)\pm0.5\Delta\log L(\HA)$].
We take account of the filter shape in the computation of the volume.
The correction can be as large as 13\% for the faint galaxies
[$L(\mathrm{H}\alpha)<10^{41}$ ergs s$^{-1}$] 
as compared to that for the brightest galaxies.
Figure \ref{Ha:LF} shows the result with the Schechter function (Schechter 1976) fit to the
$z\lesssim0.3$ \HA\ LF measured by TM98
(which is characterized by $\alpha=-1.35$, $\phi_*=10^{-2.56}$ Mpc$^{-3}$, and $L_*=10^{41.92}$
ergs s$^{-1}$; note that we convert the parameters to those of our adopted cosmology).

\begin{figure}
\epsscale{0.5}
\plotone{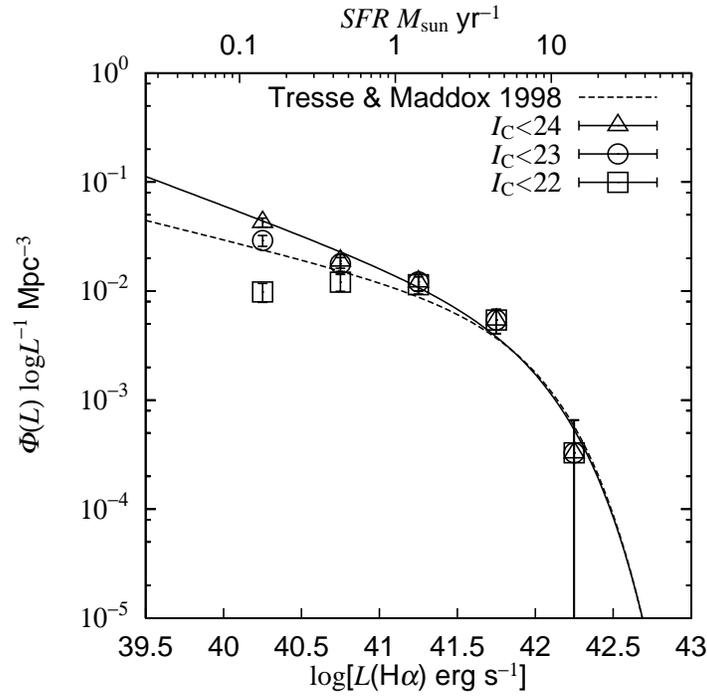}
\caption{Effects of the threshold of $I_{\rm C}$ magnitude in our \HA\ luminosity function. The solid and
dashed lines have the same meanings as those in Figure \ref{Ha:LF}.
LFs of $I_{\rm C}$-selected \HA\ emitting galaxies are shown with triangles
($I_{\rm C}<24$), circles ($I_{\rm C}<23$), and squares ($I_{\rm C}<22$).
\label{Ha:LF_Iselect}}
\end{figure}

Fitting \HA\ LF of $L(\HA)> 10^{40}$ ergs s$^{-1}$ with a Schechter function,
we find that the best-fitting parameters are 
$\alpha = -1.53 \pm 0.15$, $\log\phi_* = -2.62 \pm 0.34$, and $\log L_* = 41.95 \pm 0.25$.
We find that the faint end slope of our LF is steeper than that obtained by TM98.
This difference may be attributed to the different source selection procedures
between TM98 and ours; i.e., 
TM98 used $I$-selected CFRS galaxies lying at redshift below 0.3 with the limiting 
magnitude of $I_{\rm AB}=22.5$, while we used $\mathit{NB}816$-selected galaxies with the 
much deeper limiting magnitude of $I_{\rm C} \simeq 25$.
In order to check this possibility, we have made LFs of $I_{\rm C}<24$,
$I_{\rm C}<23$, and $I_{\rm C}<22$ sources (Figure \ref{Ha:LF_Iselect}),
using our $I_{\rm C}$-selected catalog. It is shown from this analysis that
the LF for galaxies with a brighter $I_{\rm C}$-band limiting magnitude
tend to have a  shallower faint end slope although
the brightest end of LFs is not affected by
the $I_{\rm C}$-band limiting magnitude. 
The difference between LF of TM98 and our
data is originated from this point. 
In addition to this, our survey is concentrated in a 
small redshift range around $z\approx0.24$ while the redshift range of TM98 is $z\leq0.3$.
This may also affect the LF shape.
It should be noted that our result is consistent with that obtained by
Pascual et al. (2001) although their data are available only for brighter objects
with log $L$(H$\alpha$) $\gtrsim$ 42.

\subsection{Luminosity density and star formation rate density}

The \HA\ luminosity density is obtained by integrating the LF:
\begin{equation}
	\mathcal{L}(\HA) = \int^{\infty}_0 \Phi(L)LdL = \Gamma(\alpha+2)\phi_*L_* .
\end{equation}
We then find a total \HA\ luminosity per unit comoving volume
$10^{39.65^{+0.08}_{-0.17}}$ ergs s$^{-1}$ Mpc$^{-3}$ at $z\approx0.24$ from our
best fit LF. The errors quoted here are the standard deviations taking into
account that the three Schechter parameters are correlated.
Our estimate is quite similar to that obtained by Pascual et al.
(2001), $10^{39.79\pm0.09}$ ergs s$^{-1}$ Mpc$^{-3}$.

Not all the H$\alpha$ luminosity is produced by star formation,
because active galactic nuclei (AGNs) can also contribute to the luminosity.
The amount of this contribution depends on the selection criteria.
For example, it is 8-17\% in the CFRS low-$z$ sample (Tresse et al.
1996), 8\% in the local UCM (Gallego et al. 1995), and 17-28\% in
the 15R survey (Carter et al. 2001) for the number of galaxies.
Following the manner of Pascual et al. (2001), we adopt that AGNs contribute to
15\% of the luminosity density.
We correct this AGNs contribution and then obtain the corrected luminosity
density, $10^{39.58^{+0.08}_{-0.17}}$ ergs s$^{-1}$ Mpc$^{-3}$.

\begin{figure}
\epsscale{0.5}
\plotone{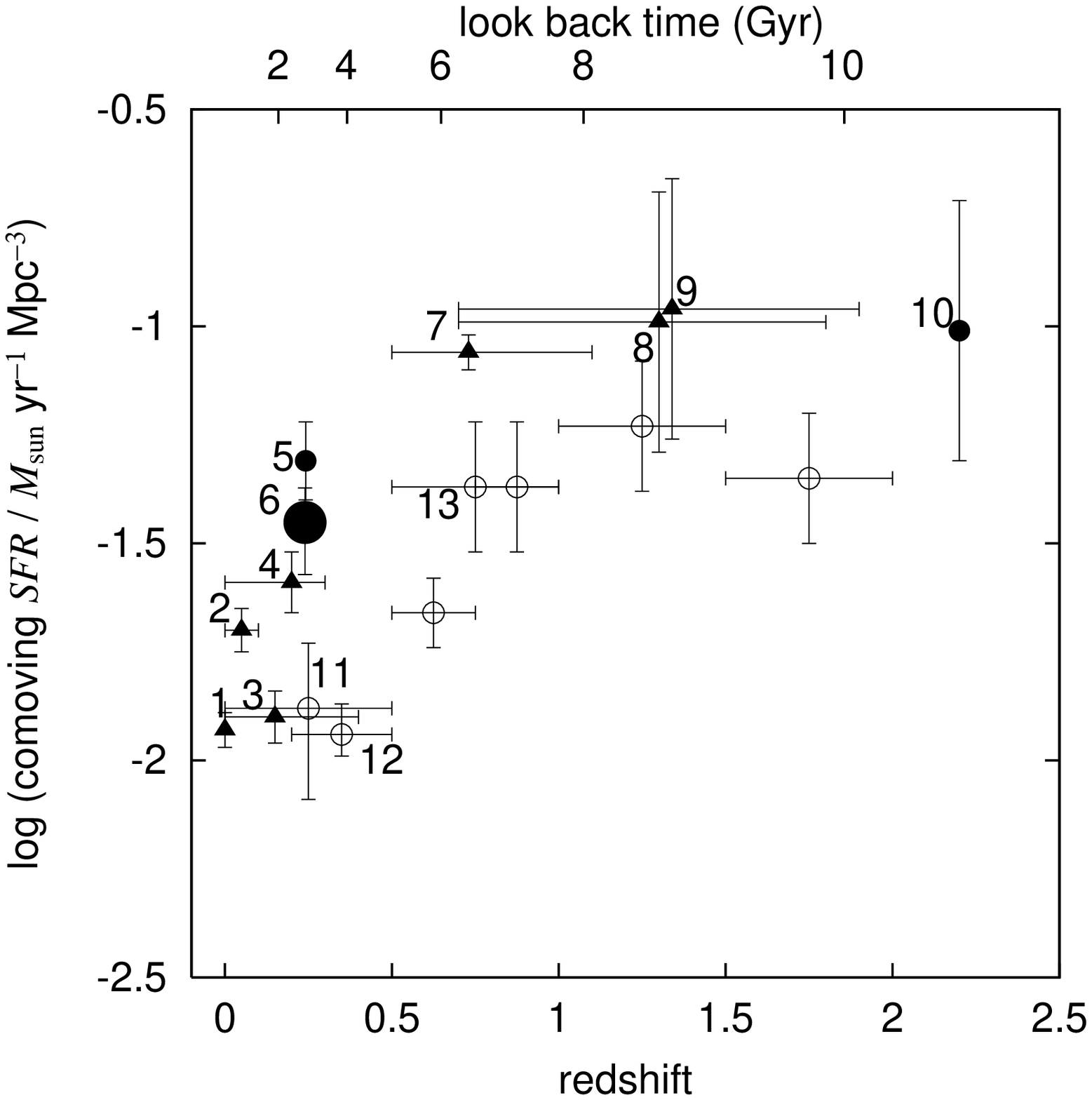}
\caption{Star formation rate density at $z\approx0.24$ derived from in this study (filled circle; ref. -- 6)
shown together with the previous investigations compiled by Trentham et al. (1999) and Tresse et al. (2002).
Filled symbols show $SFR$ derived from \HA\ LFs obtained by spectroscopy (triangles) and narrowband imaging (circles).
Open circles are $SFR$ density derived from UV LFs which includes no dust correction.
References: 1 -- Gallego et al. (1995), 2 -- Gronwall (1999), 3 -- Sullivan et al. (2000),
4 -- TM98, 5 -- Pascual et al. (2000), 7 -- Tresse et al. (2002), 8 -- Hopkins et al. (2000),
9 -- Yan et al. (1999), 10 -- Moorwood et al. (2000), 11 -- Treyer et al. (1998), 12 -- Lilly et al. (1996),
13 -- Connolly et al. (1997)
\label{Ha:MadauPlot}}
\end{figure}

The star formation rate can be estimated from the \HA\ luminosity using
$SFR = 7.9\times10^{-42}L(\HA)\:M_\Sun {\rm yr}^{-1}$,
where $L(\HA)$ is in units of ergs s$^{-1}$ (Kennicutt 1998).
Thus, the \HA\ luminosity density can be translated into the SFR density of
$\rho_{\rm SFR} \simeq 0.036^{+0.006}_{-0.012} M_\Sun$ yr$^{-1}$ Mpc$^{-3}$ (with AGN
correction $0.031^{+0.005}_{-0.010}$ $M_\Sun$ yr$^{-1}$ Mpc$^{-3}$).
Figure \ref{Ha:MadauPlot}
shows the evolution of the SFR density from $z=0$ to $z=2.0$.
The $SFR$ density measured here is higher than that of TM98. This is mainly
due to the difference of faint end slope as shown before. Our result is consistent with the
strong increase in SFR density from $z=0$ to $z\simeq1$.

We would like to thank the Subaru Telescope staff for their invaluable help. 
We thank the referee, Jesus Gallego, for his detailed comments that improved this
article. We also thank T. Hayashino for his invaluable help.
This work was financially supported in part
by the Ministry of Education, Culture, Sports, Science, and Technology (Nos. 10044052 and
10304013).

%------------------------------------------------------------------------------
%    References
%------------------------------------------------------------------------------

\clearpage

%-------------------------------------------------------------------------------
%    Table
%-------------------------------------------------------------------------------

\begin{deluxetable}{lcccc}
\tablecaption{\label{Ha:tab:cover}Emission lines potentially detected inside the
narrowband.}
\tablewidth{0pt}
\tablehead{
\colhead{Line} &
\colhead{Redshift range} &
\colhead{$\bar{z}$\tablenotemark{a}} &
\colhead{$d_L$\tablenotemark{b}} &
\colhead{$V\times10^{4}$\tablenotemark{c}} \\
\colhead{} &
\colhead{$z_1\leq z \leq z_2$} &
\colhead{} &
\colhead{(Mpc)} &
\colhead{(Mpc$^3$)}
}
\startdata
\HA\  & 0.233~~~0.251 & 0.242 & 1220  & 0.39\\
\OIII & 0.616~~~0.640 & 0.628 & 3740  & 2.39\\
\HB\  & 0.664~~~0.689 & 0.677 & 4100  & 2.62\\
\OII  & 1.17 ~~~ 1.20 & 1.19  & 8190  & 7.30\\
\enddata
\tablenotetext{a}{Mean redshift.}
\tablenotetext{b}{Luminosity distance.}
\tablenotetext{c}{Comoving volume.}
\end{deluxetable}

%-------------------------------------------------------------------------------
% figure
%-------------------------------------------------------------------------------

\begin{references}
\reference{1}{Ajiki, M., et al. 2002, ApJ, 576, L25}
\reference{1}{Barger, A. J., Cowie, L. L., \& Richards, E. A. 2000, AJ, 119, 2092}
\reference{1}{Bertin, E., \& Arnouts, S. 1996, A\&AS, 117, 393}
\reference{1}{Bruzual A., G \& Charlot, S. 1993, ApJ, 405, 538}
\reference{1}{Carte, B. J., Fabricant, D. G., Geller, M. J., Kurtz, M. J., \& McLean, B. 2001, 559, 606}
\reference{1}{Connolly, A. J., Szalay, A. S., Dickinson, M., Subbarao, M. U., \& Brunner, R. J. 1997, ApJ, 486, L11}
\reference{1}{Djorgovski, S. G., Castro, S., Stern, D., \& Mahabal, A. A. 2001, ApJ, 560, L5}
\reference{1}{Ellis, R. S., Colless, M., Broadhurst, T., Heyl, J., \& Glazebrook, K. 1996, MNRAS, 271, 781}
\reference{1}{Fan, X., et al. 2000, AJ, 120, 1167}
\reference{1}{Fujita, S. S., et al. 2003a, AJ, 125, 13}
\reference{1}{Fujita, S. S., et al. 2003b, in preperation}
\reference{1}{Gallego, J., Zamorano, J., Arag\'{o}n-Salamanca, A., \& Rego, M. 1995, ApJ, 455, L1; Erratum 1996, ApJ, 459, L43}
\reference{1}{Gallego, J., Zamorano, J., Rego, M., \& Vitores, A. G. 1997, ApJ, 475, 502}
\reference{1}{Goodrich, R. W., et al. 2001, ApJ, 561, L23}
\reference{1}{Gronwall C., 1999, in Holt S., Smith E., eds, Proc. Conf. `After the Dark Ages: When Galaxies were Young.' AIP, New York, p. 335}
\reference{1}{Iwamuro, F., et al. 2000, PASJ, 52, 73}
\reference{1}{Hopkins A. M., Connolly, A. J., \& Szalay, A. S. 2000, AJ, 120, 2843}
\reference{1}{Hogg, D. W., Cohen, J. G., Blandford, R., \& Pahre, M. A. 1998, ApJ, 504, 622}
\reference{1}{Kaifu, N. 1998, Proc. SPIE, 3352, 14}
\reference{1}{Kennicutt, R. C. 1992a, ApJ, 388,310}
\reference{1}{Kennicutt, R. C. 1992b, ApJS, 79, 255}
\reference{1}{Kennicutt, R. C. 1998 ARA\&A, 36, 189}
\reference{1}{Lilly, S. J., Le Fevre, O., Hammer, F., Crampton, D. 1996, ApJ, 460, L1}
\reference{1}{Madau, P., Ferguson, H. C., Dickinson, M. E., Giavalisco, M., Steidel, C. C., \& Fruchter, A. 1996, MNRAS, 283, 1388}
\reference{1}{Madau, P., Pozzetti, L., \& Dickinson, M. E. 1998, ApJ, 498, 106}
\reference{1}{Miyazaki, S., et al. 2002, PASJ, 54, 833}
\reference{1}{Moorwood, A. F. M.,van der Werf, P. P., Cuby, J. G., \& Oliva, E. 2000, A\&A, 362, 9}
\reference{1}{Pascual, S., Gallego, J., Arag\'{o}n-Salamanca, A., \& Zamorano, J. 2001, A\&A, 379, 798}
\reference{1}{Pettini, M., et al. 1998, in Cosmic Origins: Evolution of Galaxies, Stars, Planets, and Life, ed. J. M. Shull, C. E. Woodward, \& H. A. Thronson (San Francisco: ASP), 67}
\reference{1}{Schechter, P. 1976, ApJ, 203, 297}
\reference{1}{Shioya, Y., et al. 2002, PASJ, 54, 975}
\reference{1}{Steidel, C. C., Adelberger, K. L., Giavalisco, M., Dickinson, M., \& Pettini, M. 1999, ApJ, 519, 1}
\reference{1}{Songaila, A., Cowie, L. L., Hu, E. M., \& Gardner, J. P. 1994, ApJS, 94, 461}
\reference{1}{Sullivan, M., Treyer, M., Ellis, R. S., Bridges, B., \& Donas, J. 2000, MNRAS, 312, 442}
\reference{1}{Trentham, N., Blain, A. W., Goldader, J. 1999, MNRAS, 305, 61}
\reference{1}{Tresse, L., \& Maddox, S. 1998, ApJ, 495, 691 (TM98)}
\reference{1}{Tresse, L., Maddox, S., Le F\`{e}ver, O., \& Cuby, J.-G. 2002, MNRAS, 337, 369}
\reference{1}{Tresse, L., Maddox, S., Loveday, J., \& Singleton, C. 1999, MNRAS, 310, 262}
\reference{1}{Tresse, L., Rola, C., Hammer, F., Stasi\'{n}ska, G., Le F\`{e}vre, O., Lilly,~S.~J., \& Crampton,~D. 1996, MNRAS, 281, 847}
\reference{1}{Treyer, M. A., Ellis, R. S., Milliard, B., Donas, J., \& Bridges, T. J. 1998, MNRAS, 300, 303}
\reference{1}{Yagi, M., Kashikawa, N., Sekiguchi, M., Doi, M., Yasuda, N., Shimasaku, K., \& Okamura,~S. 2002, AJ, 123, 66}
\reference{1}{Yan, L., et al. 1999, ApJ, 519, L47}
\end{references}
\end{document}